\documentclass[12pt]{iopart}

\usepackage{iopams}

\def\be{\begin{equation}}
\def\ee{\end{equation}}
\def\ba{\begin{eqnarray}}
\def\ea{\end{eqnarray}}

\usepackage{graphicx}
\usepackage{graphicx,verbatim}
\usepackage{amsfonts}
\usepackage{amssymb}

\usepackage[colorinlistoftodos]{todonotes}
\usepackage{epstopdf}

\newcommand{\pb}[1]{\hbox{\lower0.5ex\hbox{${}_{\leftarrow}$}}\kern-1.9ex{#1}}

\def\={\,\hat{=}\,}
\def\ord{\mathcal{O}}

\def\TT{\rm TT}
\def\tt{\rm tt}
\def\t{\rm t}
\def\T{\rm T}

\def\1{(1)}
\def\2{(2)}

\def\ut#1{\rlap{\lower1ex\hbox{$\sim$}}{#1}}

\newcommand{\grad}{\nabla}

\def\f{\frac}
\def\rmd{{\textrm{d}}}
\def\ub{\underbar}

\def\vk{\vec k}
\def\vx{\vec{x}}

\def\qo{\mathring{q}}

\def\Do{\mathring{D}}

\def\vx{\vec{x}}

\def\vA{\vec{A}}
\def\vh{\vec{h}}

\def\scri{\mathcal{I}}
\def\scrip{\mathcal{I}^{+}}

\newcommand{\re}{{\rm Re}\,}
\newcommand{\im}{{\rm Im}\,}

\begin{document}

\title{
On a basic conceptual confusion in gravitational radiation theory
}
\author{Abhay Ashtekar}
\address{Institute for
Gravitation and the Cosmos \& Physics Department, Penn State, University Park, PA 16802, U.S.A.,}
\ead{ashtekar@gravity.psu.edu} 
\author{B\'eatrice Bonga}
\address{Institute for Gravitation and
the Cosmos \& Physics
  Department, Penn State, University Park, PA 16802, U.S.A.}
\ead{bpb165@psu.edu} 

\begin{abstract}

In much of the literature on linearized gravitational waves two completely different notions are called \emph{transverse traceless modes} and labelled $h_{ab}^{\TT}$, often in different sections of the same reference, without realizing the underlying inconsistency.  We compare and contrast the two notions and find that the difference persists even at leading asymptotic order near future null infinity $\scrip$. We discuss why the distinction has nonetheless remained largely unnoticed, and also point out that there are some important physical effects where only one of the notions gives the correct answer.

\end{abstract}


\maketitle

\section{The main issue}
\label{s1}

{Many of the standard references on gravitational waves begin with a decomposition of the metric perturbation $h_{ab}$ to single out its transverse traceless part $h_{ab}^{\TT}$.} (See, e.g., box 5.7 in \cite{pw}, or section 4.3 in \cite{straumann}, or section 35.4 of \cite{mtw}.) This decomposition is local in momentum space and therefore \emph{non-local in physical space}. For example, given $h_{ab}$ only in an asymptotic region, one cannot extract from it its transverse traceless part $h_{ab}^{\TT}$ even in that region. But then, while discussing gravitational waves produced by compact sources in a later section, a different construction is introduced: the transverse traceless part of $h_{ab}$ is now obtained using a projection operator $P_{a}{}^{b}$ that is \emph{local in physical space.} (See, e.g., chapter 11 of \cite{pw}, or section 4.5.1 in \cite{straumann}, or section 36.10 in \cite{mtw}.) The two notions are conceptually distinct but both are called transverse traceless and denoted by the same symbol TT, suggesting that the second notion is just a reformulation of the first. {(This point was emphasized also in \cite{racz1}.)}

Since this confusion appears to be quite common, let us begin by spelling out the difference. We will use upper case letters TT to denote the first notion and lower case letters tt to denote the second. 
In the Transverse Traceless gauge, the spatial projection $\vh_{ab}$ of the metric perturbation satisfies $\Do^{a} \big[\vh_{ab}\, -\, (1/3) (\qo^{cd}\vh_{cd})\, \qo_{ab}\big] =0$ where $\qo_{ab}$ is the spatial, Euclidean metric on the  $t={\rm const}$ slices and $\Do_{a}$ is the derivative operator it defines. Using Fourier transforms, $h_{ab}^{\TT}$ can be written as
\footnote{In this brief presentation we skip several technical points --such as the necessary fall-off conditions-- that are conceptually important but not essential for our main point. Also we do not explain the commonly used notation from the Newman-Penrose framework. These points are spelled out in the more detailed discussion of \cite{aabb2}.} 
\be \label{TT} {} \hskip-2.5cm  h_{ab}^{\TT}(t,\vx) = \f{1}{(2\pi)^{\f{3}{2}}}\,\,\int \Big(\alpha(t,\vk)\, m_{a}(\vk) {m}_{b}(\vk) + \bar\alpha(t,-\vk)\, \bar{m}_{a}(\vk) \bar{m}_{b}(\vk) \Big)\,\, e^{i\vk\cdot\vx}\, \rmd^{3}k\, , \ee
for some function $\alpha(t,\vk)$, where $\hat{k}^{a},\, m^{a}(\vk),\, \bar{m}^{a}(\vk)$ are basis vectors \emph{in momentum space} (adapted to the spherical foliation).  The second notion $h_{ab}^{\tt}$ is obtained 
by simply projecting $h_{ab}$ into the $r={\rm const}, t= {\rm const}$ 2-spheres in \emph{physical space} {using the projection operator $P_{a}{}^{b} = (m_{a}\bar{m}^{b} + \bar{m}_{a}m^{b})(\vx)$ and then removing the trace:}
\be \label{tt} {}\hskip-2.5cm h_{ab}^{\tt}(t,\vx) = \Big(P_{a}{}^{c}P_{b}{}^{d} - \f{1}{2} \, P_{ab} P^{cd}\Big)(\vx)\, h_{cd}(t,\vx) = \beta(t,\vx) m_{a}(\vx) m_{b}(\vx), \, +\, \bar\beta(t,\vx) \bar{m}_{a}(\vx) \bar{m}_{b}(\vx),\,\,  \ee 
Hence, this notion is local in physical space. Clearly the two notions have very little in common. In particular, while the spatial tensor $h_{ab}^{\tt}(t,\vx)$ has only two components, in general none of the six components of the spatial tensor $h_{ab}^{\TT}(t,\vx)$ are zero; there are four relations between them. As we will see, this is the case even to leading order in the $1/r$ expansion in the radiation zone, {where $h_{ab}^{\TT}$ captures certain physical information that escapes $h_{ab}^{\tt}$.} 

Since the main issues of the gravitational radiation theory are simpler to explain in the case of electromagnetic waves produced by spatially compact sources, monographs and review articles generally begin with Maxwell theory. We will do the same in section \ref{s2}. In section \ref{s3} we return to linearized gravity and in section \ref{s4} present the outlook. For brevity, from now on we will drop the explicit reference to $\vx$ while referring to fields in the physical space; thus for example $m_{a} \equiv m_{a}(\vx)$.

\section{$A_{a}^{\T}$ versus $A_{a}^{\t}$ in Maxwell theory}
\label{s2}

In Maxwell theory, one can work entirely with the field strength $F_{ab}$. However, in linearized gravity {one needs potentials of the perturbed Weyl tensor, e.g. to write expressions of energy-momentum and angular momentum carried away by gravitational waves.} Therefore, from the gravitational perspective  it is instructive to recast  Maxwell theory in terms of vector potentials $A_{a}$. We assume that the source current $j^{a}$ is smooth, of compact spatial support, and remains uniformly bounded in time. {(For a related discussion, see \cite{racz2}. Relation between our results and those in \cite{racz1,racz2} is discussed in \cite{aabb2}).}

The notion of Transversality on the vector potential requires $\Do^{a}\vA_{a} =0$, {where $\vA_{a}$ is the spatial projection of $A_{a}$}. Therefore, extraction of $A_{a}^{\T}$ from $A_{a}$ is a non-local operation in physical space while that of extracting $A_{a}^{\t} = P_{a}{}^{b}A_{b}$ is local. A priori the two notions are again distinct. The question is whether a simplification arises in the asymptotic $1/r$ expansion near $\scrip$. Now, there is a rich framework describing fields on $\scrip$ but it is generally tied to gauges in which the 4-potential $A_{a}$ is smooth there. While one can always go to such a gauge, it turns out that in the Coulomb gauge --which is of direct interest to the notion of Transversality-- $A_{a}$ (or its spatial projection $\vA_{a}$) {fails to be well-defined at $\scrip$.} Therefore, to compare the two notions using $\scrip$, we need to first extend the standard framework by allowing an appropriately wider class of gauges \cite{aabb2}: if we expand $A_{a}$ as
\be  
A_a = - \phi \grad_a t + \vec{A}_a  = - \phi \grad_a u + \left(-\phi +  A_1 \right)\grad_a r + A_2\, m_a + \bar{A}_2\, \bar{m}_a\, , \label{eq:vecexp}
\ee
then we have to allow gauges in which the coefficients fall off only as 
\be
\phi = \frac{\phi^{0} (u, \theta,\varphi)}{r} + \frac{\phi^{1} (u, \theta,\varphi)}{r^2} + \ldots \qquad
\hbox{and similarly for $A_{1}$ and $A_{2}$}\, . \label{falloff1}
\ee
(Field equations imply non-trivial relations between various coefficients.) To compare $A_{a}^{\T}$ with $A_{a}^{\t}$, we will need 
the behavior of various fields at $\scrip$ under this weaker fall-off. Therefore, let us make a small detour to summarize these results. 

While 4-potentials $A_{a}$ in this class can diverge at $\scrip$ in the conformally rescaled space-time --this is the case in the Coulomb gauge even for a static charge-- they do so in a well-controlled fashion. Therefore, using field equations one can establish the following results that hold in \emph{all} gauges in this class
\cite{aabb2}:\\
(i) The standard peeling properties for the Maxwell field continue to hold for the Newman-Penrose components of the Maxwell field $F_{ab}$: $\Phi_{2} = \Phi_{2}^{0}(u,\theta,\varphi)/r + \ord(1/r^{2})$;\, $\Phi_{1} = \Phi_{1}^{0}(u,\theta,\varphi)/r^{2} + \ord(1/r^{3})$;\, $\Phi_{0} = \Phi_{0}^{0}(u,\theta,\varphi)/r^{3} + \ord(1/r^{4})$. On $\scrip$, the leading order fields satisfy: $\partial_{u} \Phi_{1}^0 = \eth \Phi_{2}^{0}$ and $\partial_{u} \Phi_{0}^{0} = \eth \Phi_{1}^{0}$.\,\, Thus $\Phi_{2}^{0}$ is unconstrained on $\scrip$ and determines $\Phi_{1}^{0}$ and $\Phi_{0}^{0}$ completely, given their `initial' values at $i^{o}$.  Because of these properties, $\Phi_{2}^{0}$ is called the \emph{radiation field.} The total electric charge is a 2-sphere integral of $\re \Phi_{1}^{0}$. Thus, through its `initial' value at $i^{o}$,\, $\re \Phi_{1}^{0}$ carries the `Coulombic information' that escapes the radiation field $\Phi_{2}^{0}$.\\
(ii) The asymptotic fields $\Phi_{2}^{0}, \Phi_{1}^{0}$ and $\Phi_{0}^{0}$ are determined by the leading and sub-leading order coefficients $\phi^{0},\phi^{1},\, A_{1}^{0},  A_{1}^{1}, \, A_{2}^{0}$ in the expansion of $A_{a}$. In particular,
\be \hskip-2cm \Phi_{2}^{0} = \sqrt{2} \partial_{u} A_{2}^{0}, \qquad 
\im \Phi_{1}^{0} = \sqrt{2}\, \im \eth A_{2}^{0}, \quad {\rm and} \quad \re \Phi_{1}^{0}= \sqrt{2}\, \re \eth A_{2}^{0} + G\, , \ee
for some $G$ on $\scrip$, with $\partial_{u} G =0$. Since the complex function $A_{2}^0$ serves as the potential for $\Phi_{2}^{0}$, it is said to represent the two \emph{radiative modes} of the Maxwell field \cite{aa-bib}. Note that $A_{2}^{0}$ determines $\Phi_{2}^{0}$ and $\im \Phi_{1}^{0}$, but \emph{not} $\re \Phi_{1}^{0}$ which carries `Coulombic' information. This is just what one would expect of the radiative modes. The extra information resides in the function $G$. For example, the total electric charge is given by $Q\, = \,- (1/2\pi)\oint G(\theta,\phi)\, \rmd^{2} S$. 

We can now apply these results to the vector potential in the Transverse gauge. For notational clarity, we will use an underbar for all fields in this gauge. While in the basis vectors adapted to spherical symmetry, $\ub{A}_{a}(t,\vk) = \alpha^{\prime}(t,\vk)\, m_{a}(\vk) + \bar\alpha^{\prime}(t,\vk)\,\bar{m}_a(\vk)$ has only 2 (real) components $\alpha^{\prime}(t,\vk)$ \emph{in momentum space}, all {four} components of $\ub{A}_{a}(t,\vx)$ are non-vanishing even at $\scrip$. Furthermore, since $\ub{A}_{a}$ is gauge invariant, there is no gauge freedom in the choice of radiative modes $\ub{A}_{2}^{0}$ and $\ub{G}(\theta,\varphi)$. Next, the Transversality condition $\Do^{a} \vec{\ub{A}}_{a} =0$ implies $\partial_{u} \ub{A}_{1}^{0} = \partial_{u}\underline{\phi}^{0} =0$. Consequently while Transversality does not require any of the leading order components of $A_{a}$ to vanish on $\scrip$, it does imply that, of the four leading order components of $\ub{A}_{a}$ only two, captured by $A_{2}^{0}$, are dynamical. Expressions of the flux of energy carried away by electromagnetic waves, the electromagnetic analog of `memory', and soft charges related to infrared properties of the quantum Maxwell field, involve only the radiative modes $\ub{A}_{2}^{0}$ (see, e.g., \cite{aams,aa-bib,bg,krakow}). However, to express other physical quantities such as the total electric charge that depend also on the Coulombic aspects of the Maxwell field, one requires information in $\ub{A}_{a}$ that is \emph{not contained} in $\ub{A}_{2}^{0}$. Surprisingly, this is the case even for the local  flux $\mathcal{F}_{K}$ of angular momentum across $\scrip$, associated with a rotational Killing field $K^{a}$\, \cite{aabb2}: 
\be\mathcal{F}_{K} = T_{ab}K^{a}\tilde{n}^{b} = 2 \re \big[ (\partial_{u} \bar{A}_{2}^{0})\,\, (\sqrt{2} \eth A_{2}^{0} +G(\theta,\varphi))\,\, g(\theta,\varphi)\big] \, , \label{amflux2} \ee
where $g= K^{a}\bar{m}_{a}$. The Coulombic information, that escapes $\ub{A}_{2}^{0}$, enters through $G(\theta,\phi)$. (Same is true for the total flux, the integral of $\mathcal{F}_{K}$ on $\scrip$.) 

How does this compare to the transverse projection $A_{a}^{\t}$? Let us denote fields obtained from $A_a^{\t}$ with an under-tilde. In the physical space we now have access to only two (real) components, $\ut{A}_{2}$ of $A_{a}$. But even these are not gauge invariant. As is commonly done, let us impose the Lorenz gauge condition. Even then we have some residual gauge freedom and $\ut{A}_{2}$ fails to be invariant under it. However, its leading order (i.e. $1/r$) part $\ut{A}_{2}^{0}$ \emph{is} now gauge invariant. From our general discussion that holds in any gauge, we know that $\ut{A}_{2}^{0}$ also represents the radiative modes of the asymptotic Maxwell field. However, in general $\ut{A}_{2}^{0} \not= \ub{A}_{2}^{0}$. The difference has the form $\ut{A}_{2}^0 - \ub{A}_{2}^0 = \eth h$ where $h$ is a real function satisfying $\partial_{u} h =0$. Thus, the difference is a non-dynamical function: the wave forms obtained from $\ub{A}_{2}$ and $\ut{A}_{2}$ on $\scrip$ are simply shifted by an angle dependent function. For physical predictions, such as the expression of the energy flux across $\scrip$, that depend only on {$\partial_{u}A_{2}^{0}$}, the answers obtained using  $A_{a}^{t}$ and $A_{a}^{\T}$ agree.

However, $A_{a}^{\T}$ has much richer physical information than $A_{a}^{\t}$. In particular, the `Coulombic information' in $A_{a}^{\T}$ is simply lost in the transverse projection. In particular, if the source carries a non-zero electric charge, we \emph{cannot} obtain the correct expression of angular momentum flux using $A_{a}^{t}$ while we \emph{can} using $A_{a}^{\T}$. This is a concrete illustration of the fact that it is simply incorrect to identify the two notions of transversality.

\section{$h_{ab}^{\TT}$ versus $h_{ab}^{\tt}$ in linearized gravity}
\label{s3}

Technical calculations are much more involved for linearized gravity
but the underlying conceptual structure is completely parallel to that in Maxwell theory \cite{aabb2}. As in the Maxwell case, in order to incorporate the Transverse gauge, we need to extend the existing treatments. Much of the literature \cite{bondi-sachs,np,aa-radmodes,aa-bib} assumes (sometimes implicitly) that the conformally rescaled metric perturbation admits a smooth limit to $\scrip$. This is a consistent assumption; one can use, e.g., the Geroch-Xanthopoulos gauge \cite{gx} to satisfy this requirement. However, it is not met in the Transverse gauge. So we are led to extend the framework by allowing all gauges in which the metric components fall off only as $1/r$ in the expansion of the type (\ref{eq:vecexp}). Again, most (but not all) of the standard results continue to hold under this weaker assumption. In particular, for any metric perturbation $h_{ab}$ in this class, without assuming any gauge condition one can show that:\\
(i) The Newman-Penrose components of the linearized Weyl tensor $\Psi_{4}, \Psi_{3}$ and $\Psi_{2}$ peel properly, i.e., fall-off as $1/r,\, 1/r^{2}$ and $1/r^{3}$ respectively. As for the leading order parts in the $1/r$ expansion, given $\Psi_{4}^{0}$ on $\scrip$,\, $\Psi_{3}^{0}$ and $\Psi_{2}^{0}$ are completely determined by their  `initial' values at $i^{o}$. $\Psi_{4}^{0}$ is unconstrained and is the \emph{radiation field} at $\scrip$.\\
(ii) The leading order, i.e., $1/r$-part, of the component {$B_{22}:= h_{ab}\bar{m}^{a}\bar{m}^{b}$} of the linearized metric $h_{ab}$ now serves the role played by $A_{2}^{0}$ in Maxwell theory. In particular,
\be \Psi_{4}^{0} = - \partial_{u}^{2} B_{22}^{0}, \qquad \Psi_{3}^{0} = - \partial_{u}\,\eth {B}_{22}^{0}, \qquad \im \dot\Psi_{2}^{0} = - \im \partial_{u} \eth^{2} B_{22}^{0}\, . \ee
Hence $B_{22}^{0}$ represents the two \emph{radiative modes} of the linearized gravitational field. Furthermore, the asymptotic linearized shear $\sigma^{0}$ of the Newman-Penrose null vector $\ell^{a}$ transversal to $\scrip$ is given by $\sigma^{0} =  -(1/2\sqrt{2})\, \bar{B}_{22}^{0}$. Therefore, the linearized Bondi news tensor \cite{bondi-sachs,aa-radmodes} at $\scri$ is completely determined by $B_{22}^{0}$:
\be N_{ab} = \f{1}{\sqrt{2}}\,\,\big(\dot{B}_{22}^{0}\,m_{a}m_{b} + \dot{\bar{B}}_{22}^{0}\,\bar{m}_{a}\bar{m}_{b}\big)\, . \ee
However, again, the Coulombic information in the source is not captured by $B_{22}^{0}$. It resides in other leading order components of $h_{ab}$.

We can now compare $h_{ab}^{\TT}$ selected by the gauge condition $\Do^{a}\big[\vh_{ab} - \f{1}{3} (\qo^{cd}\vh_{cd})\,\qo_{ab}\big] =0$ with $h_{ab}^{\tt}$ selected by the projection operator $P_{a}{}^{b}$ in physical space. Again we will label fields in the TT gauge with an underbar and those obtained from the tt-projection with an undertilde. The 10 components of $\ub{h}_{ab}$ can be made gauge invariant by requiring, in addition, that the space-time component\, $\ub{h}_{cd}\,\qo^{c}{}_{a} t^{d}$\, of\, $\ub{h}_{ab}$\, be also Transverse. The two transversality conditions exhaust the gauge freedom. But they do not set any of the leading order components of $\ub{h}_{ab}$ to zero. Rather, they introduce subtle relations between them. In particular, while $\ub{B}_{22}^{0}$ can be freely specified on $\scrip$, all other 8 components are \emph{time-independent}, fixed by their values at $i^{o}$. Physical quantities that depend \emph{only} on the radiative aspects, can be expressed entirely using $\ub{B}_{22}^{0}$. These include: the flux of energy-momentum across $\scrip$ \cite{bondi-sachs,np,aams}, \, gravitational memory \cite{memory3},\, and soft charges \cite{aa-bib} that are important for the infrared behavior of the quantum gravitational field. However, with $\ub{h}_{ab}$ we also have access to other fields at $\scrip$ that capture the Coulombic aspects such as the linearized Bondi 4-momentum. The tt-projection, on the other hand, selects just two components $\ut{B}_{22}$ of $h_{ab}$ and simply discards the rest. As in the Maxwell case, $\ut{B}_{22}$ is not gauge invariant. But in the traceless and Lorenz gauge, the leading order part $\ut{B}_{22}^{0}$ at $\scrip$ \emph{is} gauge invariant and it also captures the two radiative modes of the linearized gravitational field. In general\, $\ut{B}_{22}^{0} \not= \ub{B}_{22}^{0}$\, but they are only shifted by a non-dynamical function. Nonetheless, while comparing the strain wave forms (rather than $\Psi_{4}^{0}$) obtained in the two methods, one has to bear this shift in mind because it can depend on $\ell$ and $m$ in the ${}_{s}Y_{\ell,m}$ decomposition. Finally, as in the Maxwell theory, the tt-projection would be \emph{inadequate} if a physical effect also involves the Coulombic aspect but we can use $\ub{h}_{ab}$ of the TT-method. 

\section{Outlook}
\label{s4}

In light of our discussion of the last two sections it is surprising that the distinction between $h_{ab}^{\TT}$ and $h_{ab}^{\tt}$ has been glossed over so often in the literature. The reason, we believe is two-fold. At an elementary level, the basis vectors $m_{a}({\vk})$ in momentum space (Eq. (\ref{TT})) have been implicitly identified with the $m_{a}(\vx)$ in position space (Eq. (\ref{tt})). For a TT \emph{plane wave} with wave vector $\vk_{0}$, $h_{ab}^{\TT} (t,\vx)$ has the same form as $h_{ab}^{\tt}$ \emph{if we restrict ourselves} to the two space-like directions orthogonal to $\vk_{0}$. But of course for plane waves, physical quantities such as the energy-momentum flux diverge and what is relevant physically are superpositions, as in (\ref{TT}). And for these, it is incorrect to ignore the distinction. The second and more subtle reason is brought out by our analysis: Even though the two methods are completely unrelated, there is agreement between their final expressions of physical quantities that involve \emph{only} the radiative modes. However there are also physical quantities requiring Coulombic information that can be computed in the TT gauge but not by using the tt-projection. In addition to the quintessentially Coulombic quantities such as the electric charge or Bondi mass, other physically interesting quantities can involve a subtle interplay between the radiative and Coulombic modes, as the example of angular momentum of section \ref{s2} shows. If we consider an oscillating dipole (with zero total charge), one finds that $G(\theta,\phi)=0$ and angular momentum flux can be expressed entirely in terms of $A_{2}^{0}$. But if we superimpose on it the Coulomb field of a static electric charge, then $G(\theta,\phi)\not=0$ and angular momentum can no longer be expressed using just the radiative modes. In the gravitational case, the source carrying a time changing quadrupole moment will also have a non-zero mass and therefore we expect that the radiative modes will not suffice to determine the angular momentum flux; as in the Maxwell case, the tt-method will give incorrect results. 

Since there is no gauge invariant stress energy tensor for gravitational waves, to calculate fluxes across $\scrip$, one needs to adopt an alternate strategy. For energy-momentum we already have one \cite{aabb2} but for angular momentum we would have to extend the existing phase space of radiative modes \cite{aams} to include the appropriate Coulombic information. This step has been carried out for the Maxwell field and the extended phase space does lead to the correct angular momentum flux even though the calculation does not refer to the stress-energy tensor at all. Therefore one can hope to extend it to the gravitational case. The general method entails an extension of the phase space of radiative modes at $\scrip$ to accommodate the presence of sources --such as compact objects-- without including them explicitly in the phase space. In the Maxwell case, this is possible by including information about the Coulomb field of sources, without having to introduce new degrees of freedom to describe the sources themselves. The issue is being analyzed in the gravitational case both at the linear and non-linear level.
It is interesting that results obtained while investigating the relation between $h_{ab}^{\TT}$ and $h_{ab}^{\tt}$ have opened this unforeseen window.

The tt-projection fits nicely with Bondi-Sachs type expansions in $1/r$. This is a definite advantage in the asymptotically flat context where the $\TT$-method is cumbersome to use. However, the other side of the coin is that it seems impossible to extract gauge invariant physics from the tt-method in other contexts. The TT-method is free from this drawback. This is why one finds no mention of the tt-projection in the work on primordial gravitational waves where one exclusively uses $h_{ab}^{\TT}$. So far, however, mainstream cosmological work does not include the study of retarded gravitational waves produced by compact bodies. As gravitational wave detectors extend their reach --especially through space-based detectors-- the interface between astrophysics and cosmology will become increasingly important. We will have to develop viable theoretical frameworks to describe gravitational radiation from compact sources that are embedded in our cosmological --rather than asymptotically Minkowskian-- space-time. Then it is likely that the Bondi-Sachs framework and the approximation methods tailored to it will no longer be {directly} useful. Indeed, already in the study of linearized gravitational waves on the de Sitter background, one has to develop entirely new techniques \cite{abk-prl}. Can one develop approximation methods that are as `user-friendly' as the tt-projection but viable also in this more general context?

Detailed derivations, the precise underlying assumptions, and additional results can be found in \cite{aabb2}.  

\section*{Acknowledgment} We thank Eric Poisson and especially Aruna Kesavan for discussions during the early part of this work. {Thanks are also due to Badri Krishnan who drew our attention to \cite{racz1} after learning of our main results, and Istav\'{a}n I.~R\'{a}cz  who informed us about \cite{racz2} after our article appeared on the arXiv.} This work was supported in part by the NSF grant PHY-1505411, the Eberly research funds of Penn State and Mebus Graduate Fellowship to BB. AA thanks the Erwin Schr\"odinger Institute, Vienna for hospitality.

\end{document}